%% file: main.tex
\documentclass{article}

\PassOptionsToPackage{numbers}{natbib}
\usepackage[preprint]{neurips_2024}

\usepackage[utf8]{inputenc} 
\usepackage[T1]{fontenc}    
\usepackage{url}            
\usepackage{booktabs}       
\usepackage{amsfonts}       
\usepackage{nicefrac}       
\usepackage{microtype}      

\usepackage[dvipsnames, table]{xcolor}

\usepackage{url}            
\usepackage{booktabs}       
\usepackage{amsfonts}       
\usepackage{nicefrac}       
\usepackage{microtype}      
\usepackage{bbm}
\usepackage{amsmath}
\usepackage{tikz}
\usepackage{adjustbox}
\usepackage{enumerate}
\usepackage{fontawesome}
\usepackage{amssymb}
\usepackage{subcaption} 
\usepackage{wrapfig}
\usepackage{multirow}
\usepackage{graphicx}
\usepackage{soul}
\usepackage[most]{tcolorbox} 
\usepackage{verbatim}
\usepackage{wrapfig}
\usepackage{bbding}

\newcommand{\hlbox}[2][yellow]{%
  \begingroup
  \sethlcolor{#1}%
  \fbox{\hl{#2}}%
  \endgroup
}
\newtcolorbox{block}[1]{%
  colback=emobotyellow!10!white,
  colframe=emobotyellow!70,
  coltitle=black, 
  title=#1
}

\usetikzlibrary{decorations}
\usetikzlibrary{decorations.pathmorphing}
\usetikzlibrary{decorations.pathreplacing}
\usetikzlibrary{decorations.shapes}
\usetikzlibrary{decorations.text}
\usetikzlibrary{decorations.markings}
\usetikzlibrary{decorations.fractals}
\usetikzlibrary{decorations.footprints}
\usetikzlibrary{arrows.meta}

\definecolor{burgundy}{rgb}{0.5, 0.0, 0.13}
\definecolor{blueish}{RGB}{176,224,230}
\definecolor{emobotyellow}{HTML}{F5C24A}
\definecolor{C4}{RGB}{0,148,181}
\definecolor{B4}{RGB}{14,135,201}
\definecolor{myblue}{HTML}{0064E0}

\newcommand\myshade{90}

\colorlet{mylinkcolor}{B4}
\colorlet{mycitecolor}{B4}
\colorlet{myurlcolor}{C4}

\usepackage[citecolor=mycitecolor!\myshade!black,
		      colorlinks=true,
		      linkcolor=myblue!\myshade!black
            ]{hyperref}

\newcommand{\x}{\mathbf{x}}
\newcommand{\h}{\mathbf{h}}
\newcommand{\z}{\mathbf{z}}
\newcommand{\p}{\mathbf{p}}
\newcommand{\y}{\mathbf{y}}
\usepackage{xspace}
\newcommand{\emobot}{\raisebox{-0.1\height}{\includegraphics[scale=0.022, trim=100 80 100 70, clip]{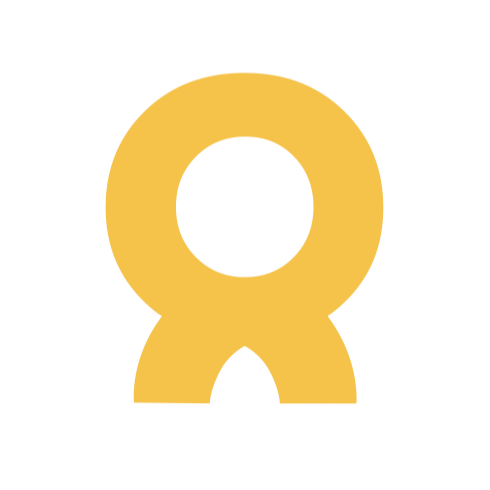}}}
\newcommand{\LargeEmobot}{\raisebox{-0.1\height}{%
            \includegraphics[scale=0.038, trim=100 80 100 70, clip]{
            img/emobot_logo.png}
        }
    }
\newcommand{\EmoLLM}{Em\protect\emobot SLLM\xspace}
\newcommand{\EmoLLMbf}{\textbf{Em}\protect\emobot\textbf{SLLM}\xspace}

\title{Em\LargeEmobot \hspace{-0.3cm}SLLM: Parameter-Efficient Adaptation of LLMs for Speech Emotion Recognition}

\author{%
  Hugo Thimonier\textsuperscript{1,*}
  \quad Antony Perzo\textsuperscript{1}
  \quad Renaud Seguier\textsuperscript{1, 2} \\
  \textsuperscript{1} Emobot\emobot, Paris, France \\ \texttt{ \{name\}.\{surname\}@emobot.fr } \\
  \textsuperscript{2} CentraleSup\'elec, IETR (UMR CNRS 6164), Rennes, France\\
  \textsuperscript{*} Corresponding author.
}

\begin{document}

\maketitle

\begin{abstract}
   Emotion recognition from speech is a challenging task that requires capturing both linguistic and paralinguistic cues, with critical applications in human-computer interaction and mental health monitoring. Recent works have highlighted the ability of Large Language Models (LLMs) to perform tasks outside of the sole natural language area. In particular, recent approaches have investigated coupling LLMs with other data modalities by using pre-trained backbones and different fusion mechanisms. This work proposes a novel approach that fine-tunes an LLM with audio and text representations for emotion prediction. Our method first extracts audio features using an audio feature extractor, which are then mapped into the LLM's representation space via a learnable interfacing module. The LLM takes as input (1) the transformed audio features, (2) additional features in the form of natural language (e.g., the transcript), and (3) a textual prompt describing the emotion prediction task. To efficiently adapt the LLM to this multimodal task, we employ Low-Rank Adaptation (LoRA), enabling parameter-efficient fine-tuning. Experimental results on standard emotion recognition benchmarks demonstrate that our model outperforms all but one existing Speech-Text LLMs in the literature, while requiring less than half the parameters of competing approaches. This highlights our approach's effectiveness in integrating multi-modal inputs for speech-based emotion understanding while maintaining significant computational efficiency.
\end{abstract}

\input{1_introduction}
\input{2_related_works}
\input{3_method}
\input{4_experiments}
\input{5_discussion}
\input{6_conclusion}

\bibliographystyle{plainnat}
\bibliography{bibliography}
\newpage 
\appendix

\input{appendix}

\end{document}

%% file: 1_introduction.tex
\section{Introduction}
\label{sec:introduction}

    Predicting the emotion conveyed in audio is a critical task with many healthcare applications. For instance, tracking a patient’s emotional fluctuations throughout the day can offer psychiatrists valuable insights into conditions such as depression—a disorder characterized by persistent sadness, irritability, and apathy \cite{depression_children}. As a result, continuous and non-invasive emotion monitoring could significantly improve diagnostic accuracy and treatment personalization.

    The widespread adoption of smartphones among both minors and adults \cite{fischer2019risk, smartphoneuse} has enabled scalable, real-time monitoring of behavioral and emotional health. Among the modalities accessible through smartphones, speech is particularly informative due to its rich linguistic and paralinguistic content \cite{PISANSKI2016304, Ding2023}. These cues have been linked to various mental health conditions, and numerous studies have explored speech emotion recognition (SER) as a proxy for psychological well-being \cite{Yang_2016, Hansen_2022}.
    
    SER has been addressed lately by
    leveraging feature representations coming from models trained for different tasks \cite{chung2019unsupervised, Liu2020NonAutoregressivePC, Liu2020TERASL, Baevski2020vq-wav2vec, Chen2021WavLMLS, Baevski2020wav2vec2A, hubert2021}. For instance, \cite{WangWav2vecSER} fine-tune HuBERT \cite{hubert2021} and Wav2Vec2.0 \cite{Baevski2020wav2vec2A} for the task of SER \cite{Wagner_2023}. Other approaches consider the use of frozen self-supervised models as feature extractors to train a supervised classifier \cite{pepino21_interspeech, emobot_speech2024} by solely adding a linear layer on top of the self-supervised model. While promising, these approaches are quite simple and often rely exclusively on speech-related information.
    
    Given the recent discoveries on the strong capacities of LLMs for multimodal tasks, research has been oriented towards leveraging LLMs for other modalities, including audio. In particular, different overlapping lines of works have been considered: LLMs that \textit{speak}, LLMs that \textit{listen}, and LLMs that can do both. Relevant to the present work is LLMs that \textit{listen}, which describe LLMs that can take as input both natural language and audio features \cite{wu2023decoderonlyarchitecturespeechtotextlarge, hu-etal-2024-wavllm, kong2024audio, Das2024SpeechVerseAL, tang2024salmonn, chu2024qwen2audiotechnicalreport, pandey2025sift50mlargescalemultilingualdataset}.
    
    Current state-of-the-art LLM-based approaches for Speech Emotion Recognition, such as SIFT-LLM \cite{pandey2025sift50mlargescalemultilingualdataset} and SALOMONN \cite{tang2024salmonn}, demonstrate impressive performance but rely on models with over 7 billion parameters. This makes them impractical for privacy-sensitive, on-device deployment—an essential consideration when handling highly personal data like a user’s emotional state over time.
    
    In the present work, we propose a parameter-efficient approach LLM-based approach for speech emotion recognition. We build on \cite{vallaeys2024improvedbaselinesdataefficientperceptual} and use as a downsampling module an attention-based model that selects learnable queries to represent the audios to be fed to the LLM. We rely on their feature mapping mechanism, QPMapper, as it is lightweight and has shown strong performance for visual and audio data inclusion in LLMs. We rely on Robust wav2vec 2.0 \cite{hsu21_interspeech} and WavLM as the audio feature extractors and experiment using Llama3.2-3B-Instruct \cite{grattafiori2024llama3herdmodels}. We train our model using a 3-step learning curriculum. In the first phase, we treat automatic speech recognition (ASR) as a proxy task to align the audio representations with the LLM embedding space. During this phase, the audio encoder and LLM are frozen, and only the QPMapper is updated. In the second phase, we continue training on the ASR task but enable fine-tuning of the LLM via Low-Rank Adaptation (LoRA) \cite{hu2022lora}, allowing the language model to begin adapting to audio-conditioned tasks. Finally, in the third phase, we introduce the SER task to specialize the model for emotion recognition, further fine-tuning the LLM with LoRA while continuing to update the weights of the downsampling module.
    
    We compare our model, coined \textbf{Emo}tion \textbf{S}peech \textbf{L}arge \textbf{L}anguage \textbf{M}odel (\EmoLLM), to existing text-audio language models for the task of speech emotion recognition. \EmoLLM achieves competitive SER performance, outperforming all but one existing text-audio model while maintaining a substantially smaller parameter footprint. This demonstrates its potential for privacy-preserving, on-device emotion recognition. In addition, we carefully design prompts to guide the language model's reasoning over the audio representations, which we find to be essential for improving emotion recognition accuracy in low-resource settings.
    
    The remainder of the paper is organized as follows: Section \ref{sec:related_works} presents the related works on self-supervised learning and multimodal LLMs. Section \ref{sec:method} presents the approach used to address the SER problem by using LLMs. Section \ref{sec:results} shows the experimental results of our proposed methodology. In Section \ref{sec:discussion} we perform some ablations to assess the relevance of our approaches key characteristics. Finally, in \ref{sec:conclusion} we conclude and discuss the limitations of the present work.

%% file: 2_related_works.tex
\section{Related Works}
\label{sec:related_works}

    We review three key areas of prior work that underpin our approach: self-supervised learning for audio representations, the emergence of multimodal large language models, and recent advances in combining speech and text within LLM frameworks.

    \paragraph{Self-Supervised Learning}
        Self-supervised learning (SSL) for representation learning consists in solving a well-chosen pretext task to generate a meaningful representation of the data that can serve for downstream tasks. It has garnered increasing attention in recent years due to its success in fostering performance on downstream tasks on different data modalities. In the case of visual modality, approaches have been developed for images and video sequences. For images methods such as I-JEPA \cite{Assran_2023_CVPR}, SwAV \cite{caron2020unsupervised}, VICReg \cite{bardes2022vicreg} or Barlow Twins \cite{zbontar2021barlow} have been explored. In the case of video sequences, approaches such as V-JEPA \cite{bardes2024revisiting}, LatentMIM \cite{latentMIM} or GTCC \cite{Donahue_2024_CVPR} have been studied. These learned representations have been shown to improve the performance of supervised downstream tasks.
        Similarly, methods for tabular data modality such as XTab \cite{xtab2023}  BinRecon \cite{pmlr-v235-lee24v}, SwitchTab \cite{switchtab_2024}, or T-JEPA \cite{thimonier2025tjepa} enable better training of deep methods on tabular data. 
        For audio data, self-supervised methods have also emerged as a powerful approach to learning representations without labeled supervision. Some seminal approaches include Wav2Vec \cite{schneiderwav2vec}, which was later improved with Wav2Vec2.0 \cite{Baevski2020wav2vec2A}. The latter approach involves mapping the original audio signal to a latent space with a convolutional model, which is then mapped to a context space using a transformer model. Their approach also includes a quantization module following previous work \cite{vqvae, vqvae2, Baevski2020vq-wav2vec}. This quantization involves masking part of the audio representation and selecting the masked elements among some distractors. Another seminal work includes HuBERT \cite{hubert2021}, in which the self-supervised task predicts hidden cluster assignments of the masked frames. In addition to this approaches, WavLM \cite{Chen2021WavLMLS} involves a transformer model and the mask reconstruction task and displays strong performance; BYOL-A \cite{byol-a, byola2} that adapts the BYOL approach \cite{BYOL} originally proposed for images to audio.
        While subsequent works have proposed various enhancements, Wav2Vec2.0, HuBERT, and WavLM continue to dominate as the primary self-supervised models for audio.

    \paragraph{Multimodal Large Language Modeling}
        Prior works have tried to leverage LLM to include modalities other than natural language. The main emphasis has been on including images in LLMs, coined as Vision Language Models (VLMs). VLMs could be described as LLMs that are also able to receive images as inputs. \cite{bordes2024introductionvisionlanguagemodeling} distinguish several (possibly overlapping) approaches to VLMs: mask-based approaches \cite{Singh_2022_CVPR, kwon2023masked}, contrastive-based approaches \cite{clip2021, Lliplavoie24a}, generative approaches \cite{chameleonteam2024chameleonmixedmodalearlyfusionfoundation, yu2022coca}, and approaches based on pretrained backbones \cite{tsimpoukelli2021multimodal, zhu2023minigpt4enhancingvisionlanguageunderstanding}. 
    
    \paragraph{Speech-Text LLMs}
        Efforts have also been put on enabling LLMs to handle audio and speech modalities. Due to their lighter training cost, most approaches have been oriented toward including pretrained backbones to fuse modalities in LLMs rather than training models from scratch. The first approaches focused on specific tasks and enabled LLMs to perform one audio task at a time, such as ASR \cite{Fathullah2024, Yu2024}. Later work has focused on multi-task models that can perform several tasks. For instance, \cite{tang2024salmonn} propose SALMONN in which they combine Whisper \cite{pmlr-v202-radford23a} and BEATS \cite{pmlr-v202-chen23ag} with an LLM by using a window-level Q-Former \cite{blip2023}, and train their model to perform tasks ranging from ASR, music captioning or emotion recognition. Similarly, Qwen2-audio \cite{chu2024qwen2audiotechnicalreport} also relies on Whisper as the audio encoder. It combines the audio encoder's output representation with the tokenized representation of the text before feeding it to an LLM. They then proceed to train the LLM on more than 30 tasks, including ASR, SER, Speech-to-Text Translation or Vocal Sound Classification.
        Other examples of multi-task approaches include methods such as Speechverse \cite{Das2024SpeechVerseAL}, which relies on both WavLM \cite{Chen2021WavLMLS} and Flan-T5 \cite{chung2022scalinginstructionfinetunedlanguagemodels}; or WavLLM \cite{hu-etal-2024-wavllm} that relies on WavLM and Llama-2 7B-Instruct as the backbone models.

%% file: 3_method.tex
\section{Method}
\label{sec:method}

    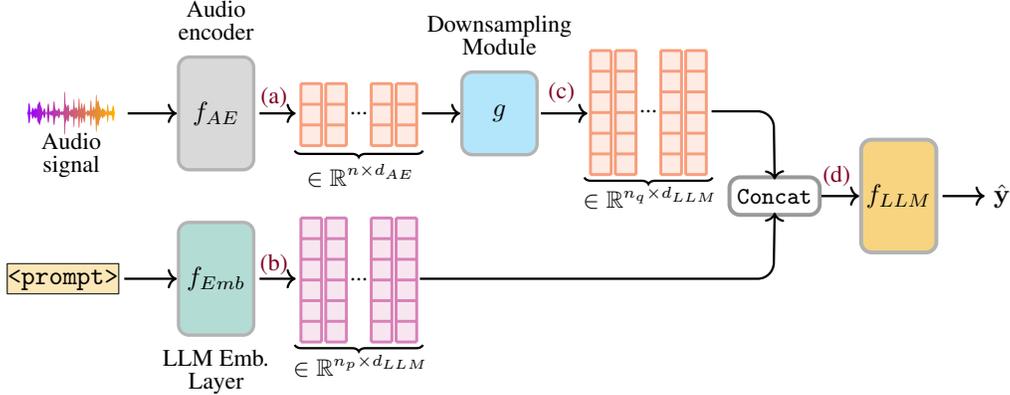
\begin{figure*}[t!]
        \centering
        \input{img/img_EmoLLM}
        \caption{\EmoLLMbf \textbf{Pipeline}. In step {\color{burgundy}(a)}, the audio signal is fed to a pretrained audio encoder to obtain a vectorized embedded representation $\h_{AE}$ of dimension $\mathbb{R}^{n \times d_{AE}}$. In step {\color{burgundy}(b)} a text prompt is fed to the embedding module of an LLM to output a vectorized embedded representation $\mathbf{p}$ of dimension $\mathbb{R}^{n_p \times d_{LLM}}$. In step {\color{burgundy}(c)} $\h_{AE}$ is fed to a downsampling module and the obtained sequence $\h_{ds}$ is of dimension $\mathbb{R}^{n_q \times d_{LLM}}$. In step {\color{burgundy}(d)} $\h_{ds}$ and $\mathbf{p}$ are concatenated and fed to the LLM, $f_{LLM}$ that predicts the target given a task (ASR or SER).}
        \label{fig:EmoLLM}
    \end{figure*}
    We put forward \EmoLLM, an LLM that integrates both audio and textual modalities to enhance emotion prediction by fine-tuning a large language model (LLM) with Low-Rank Adaptation (LoRA) (see Figure \ref{fig:EmoLLM}). We first extract audio features using a pretrained audio encoder, then project them into the LLM’s representation space using a downsampling module. The resulting audio representation is concatenated with the embedded instruction prompt, and the combined input is passed to the LLM for causal generation.
    
    \subsection{Architecture}
    \label{subsec:architecture}

        \paragraph{Audio encoder} 
            To extract semantically useful features from an audio signal, we rely on a pretrained audio feature extractor. Let $\x$ denote an audio signal, and $f_{AE}$ denote the audio feature extractor.
            \begin{equation}
                \label{eq:audio_encoding}
                \h_{AE} = f_{AE}(\x; \theta_{AE}) \in \mathbb{R}^{n \times d_{AE}},
            \end{equation}
            where $d_{AE}$ is the hidden dimension of the audio encoder, and $n$ the sequence length of the model's output.
        
        \paragraph{Downsampling module} 
            As highlighted in prior work \cite{Das2024SpeechVerseAL, défossez2024moshispeechtextfoundationmodel}, audio signals typically yield much less compact latent representations compared to other modalities such as natural language. For example, the tokenized representation of a spoken sentence, when encoded by an audio foundation model, results in a substantially longer sequence than the corresponding textual representation. Consequently, directly concatenating the audio and text embeddings would introduce a disproportionate bias toward the audio modality, potentially skewing downstream tasks.
            To alleviate this issue, we downsample the audio representation $h_{AE}$ obtained in Eq. \eqref{eq:audio_encoding} to shorten its sequence length $n$. 

            We adopt a Query Pooling Mapper (QPMapper) module, previously shown to perform well on image modalities \cite{vallaeys2024improvedbaselinesdataefficientperceptual}. In a nutshell, this module adds $n_{q}$ learnable queries, $\mathbf{q} \in \mathbb{R}^{n_q \times d_{AE}}$, to the original sequence $\h_{AE}$. This concatenated sequence is then passed through a transformer encoder, and the output queries' representations are kept as the downsampled audio representation. Additionally, this downsampling module $g$ serves to project the audio features into the dimensional space of the language model’s representations. Thus,
            \[
            \h_{ds} = g(\h_{AE}; \theta_{ds}) \in \mathbb{R}^{n_q \times d_{LLM}},
            \]
            where $n_q$ is a hyperparameter.

        \paragraph{Large language model} 
            Let $f_{LLM}$ denote the large language model that will serve for the causal generation. It inputs a concatenated sequence comprised of: (i) the output of the downsampling module, $\h_{ds} \in \mathbb{R}^{n_q \times d_{LLM}}$, (ii) an embedded vectorized text prompt describing the task $\mathbf{p} \in \mathbb{R}^{n_p \times d_{LLM}}$ and (iii) possibly some textual information extracted from the audio signal also embedded and vectorized, e.g. a transcript, $\z \in \mathbb{R}^{n_z \times d_{LLM}}$. We further discuss in section \ref{subsec:add_features} the information contained in $\z$ and its impact on the overall performance.
            Thus, the probability distribution over the output from \EmoLLM can be expressed as:
            \begin{equation}
                \text{\EmoLLM}(\x, \mathbf{p}, \z) = f_{LLM}\left ( [\h_{ds}, \mathbf{p}, \z]; \theta_{LLM} \right).
            \end{equation}

        \subsection{Optimization objective}

            For each task, we provide a set of 10 prompts, selected randomly for each sample in a batch. We display in sections \ref{app:subsec:asr} and \ref{app:subsec:ser} examples of each prompt for both SER and ASR. For each task $t \in [\text{ASR}, \text{SER}]$, we train the LLM for next-token prediction in an auto-regressive manner as in standard LLM training schedules.
            
            Formally, let $\mathbf{p}^t$ denote a prompt for task $t$ uniformly sampled on $\mathcal{P}^{t}$ the set of prompts for task $t$, and $\mathcal{D}^t$ the set of datasets used for task $t$. Each sample can be represented as a tuple $(\x^t, \mathbf{p}^t, \z^t, \mathbf{y}^t)$, where $\x^t$ is the audio waveform, $\mathbf{p}^t$ the sampled prompt for the corresponding task, $\z^t$ some additional information relevant for task $t$ and $\mathbf{y}^t$ the label to be predicted. The probability of predicting the label $\y^t$ is modeled as
            \begin{equation}
                \label{eq:EmoLLM}
                p(\y^t \mid \x^t, \p^t, \z^t; \Theta) = \text{\EmoLLM}(\x^t, \p^t, \z^t),
            \end{equation}
            where $\Theta = \{\theta_{AE}, \theta_{ds}, \theta_{LLM}\}$. The LLM can attend to all tokens in the concatenated sequence $[\x^t, \p^t, \z^t]$ and is trained to leverage the audio tokens to minimize the negative likelihood given the probability modeled in Eq. \eqref{eq:EmoLLM}. The negative likelihood for target $\y^t$ for each sample in the training set is thus expressed as
            \begin{equation}
                \mathcal{L}(\Theta) = - \log p (\y^t \mid \x^t, \p^t, \z^t; \Theta),
            \end{equation}
            and is minimized using gradient descent.

    \subsection{Data concatenation}
        \label{subsec:data_concat}

        We observed experimentally that \textit{directly} concatenating the downsampled audio representation $\h_{ds}$ with the textual prompt tokens $\p$ that guides the LLM, i.e. forming the input as $[\h_{ds},\p]$, can hinder the LLM’s ability to effectively attend to the audio tokens during training. To address this, we prepend a natural language cue, \hlbox[emobotyellow!40]{\texttt{Here are some audio tokens:}}, to the audio token sequence. This guiding phrase helps the model identify and contextualize the incoming audio tokens. An alternative strategy would be to enclose the audio tokens within special markers, such as \hlbox[emobotyellow!40]{\texttt{<|audio|>}, \texttt{<|/audio|>}}; however, this approach would necessitate modifying the LLM’s embedding matrix, thereby introducing significant additional computational cost.

    \subsection{Task prompt}
        \label{subsec:task_prompt}
        
        \paragraph{Prompt selection} 
            When prompting the LLM, we carefully describe the task to be performed and construct a pool of 20 distinct prompts per task. During training, each sample is randomly assigned a prompt uniformly sampled from the corresponding task-specific pool. For the Speech Emotion Recognition (SER) task, we use a closed-form prompt that explicitly enumerates the emotion categories from which the model should select its answer. To reduce positional bias and encourage generalization, the order of these categories is randomized for each sample. Empirically, we found that this randomization helps mitigate overfitting during training.

        \paragraph{Prompt format} 
             We explicitly detail in the system prompt the expected format of the answer that the LLM should provide. Following previous work \cite{Das2024SpeechVerseAL}, for single task prediction, we ask for the answer to be formatted as  \hlbox[emobotyellow!40]{\texttt{| {TASK}: <answer> |}} where \texttt{TASK} can either be \texttt{ASR} or \texttt{Emotion}. For joint prediction, latter discussed in section \ref{subsec:joint_prediction}, we ask for a similar formatting of the answer, in the format \hlbox[emobotyellow!40]{\texttt{| {ASR}: <answer> | Emotion: <answer> |}}.

    \subsection{Training curriculum}
        \label{subsec:training_schedule}
    
        We rely on a three-stage curriculum learning framework. First, we train the model only on the ASR task following previous work \cite{hu-etal-2024-wavllm, Das2024SpeechVerseAL}. In this first phase (\textcolor{emobotyellow}{\bf P1}), the audio feature extractor and LLM are frozen, while the downsampling module's weights are the only components updated. This phase aims to learn an effective mapping from the audio representation space to the LLM’s embedding space, allowing the model to leverage the LLM’s semantic capabilities \cite{Das2024SpeechVerseAL, pandey2025sift50mlargescalemultilingualdataset}. In the second phase (\textcolor{emobotyellow}{\bf P2}), we introduce Low-Rank Adaptation (LoRA) adapters to the LLM and continue training on the ASR task. Here, both the downsampling module and the LLM (via LoRA) are fine-tuned jointly, enabling the model to begin adapting to audio-conditioned language tasks. Finally, in the last phase (\textcolor{emobotyellow}{\bf P3}) we introduce the SER objective and train the model to perform both ASR and emotion recognition simultaneously.

%% file: img/img_emoLLM.tex
\begin{tikzpicture}[scale=0.55]

    \node (audio) at (-1, 3) {\includegraphics[width=1.25cm]{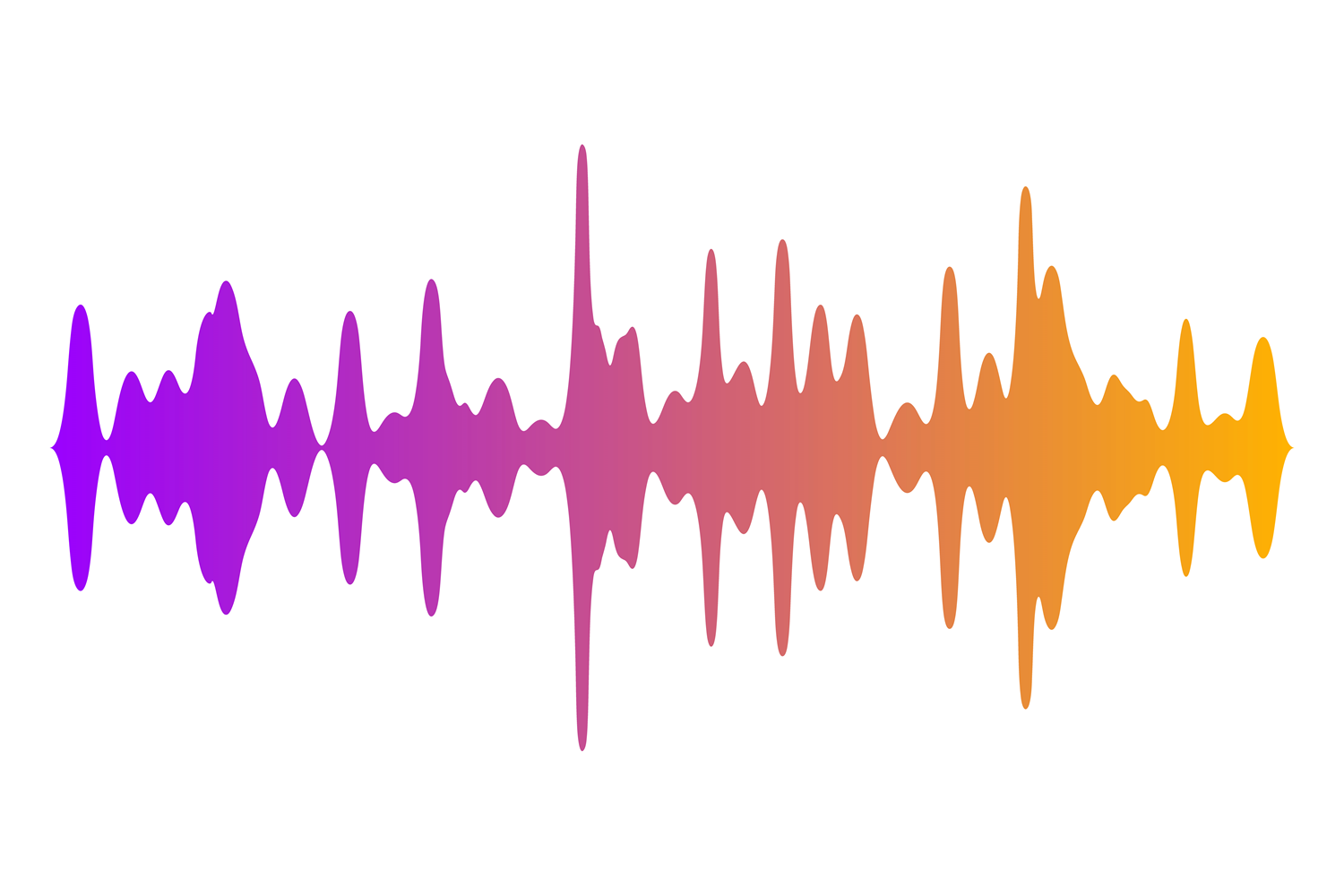}};
    \node[align=center] () at (-1,2) {\small{Audio} \\[-0.5ex] \small{signal}};

    
    \node (prompt) at (-1.2, -1) {\hlbox[emobotyellow!40]{\texttt{<prompt>}}};
    

    \node[align=center] () at (2.5,5.25) {\small{Audio} \\[-0.5ex] \small{encoder}};

    \node[rectangle, draw, fill=gray, minimum width=1.cm, minimum height=1.5cm, align=center, rounded corners, opacity=0.3, line width=1.25pt,] (f_AE) at (2.5,3.) {};
    \node () at (2.5, 3.) {$f_{AE}$};

    \node[rectangle, draw, fill=PineGreen, minimum width=1.cm, minimum height=1.5cm, align=center, rounded corners, opacity=0.3, line width=1.25pt] (f_emb) at (2.5,-1.) {};
    \node () at (2.5, -1.) {$f_{Emb}$};

    \node[align=center] () at (2.5,-3.25) {\small{LLM Emb.} \\[-0.5ex] \small{Layer}};

    
    \begin{scope}[yshift=-0.25cm, name=Melon_sample_z_left]
        \draw[Melon, step=0.5cm, thick, xshift=0.05cm, yshift=-0.03cm, line width=1.1pt] (4.5-0.001,4.+0.001) grid (5-0.0001,2.5-0.001);
        \path [anchor = center, fill=Melon, opacity=0.2, xshift=0.05cm, yshift=-0.03cm] (4.5,4) rectangle (5,2.5);
    \end{scope}

    \begin{scope}[yshift=-0.25cm, xshift=0.1cm, name=Melon_sample_z_left]
        \draw[Melon, step=0.5cm, thick, xshift=0.05cm, yshift=-0.03cm, line width=1.1pt] (5-0.001,4.+0.001) grid (5.5-0.0001,2.5-0.001);
        \path [anchor = center, fill=Melon, opacity=0.2, xshift=0.05cm, yshift=-0.03cm] (5,4) rectangle (5.5,2.5);
    \end{scope}

    \begin{scope}[yshift=-0.25cm, xshift=0.7cm, name=Melon_sample_z_right]
        \draw[Melon, step=0.5cm, thick, xshift=0.05cm, yshift=-0.03cm, line width=1.1pt] (5.5-0.001,4+0.001) grid (6-0.0001,2.5-0.001);
        \path [anchor = center, fill=Melon, opacity=0.2, xshift=0.05cm, yshift=-0.03cm] (5.5,4) rectangle (6,2.5);
    \end{scope}

    \begin{scope}[yshift=-0.25cm, xshift=0.8cm, name=Melon_sample_z_right]
        \draw[Melon, step=0.5cm, thick, xshift=0.05cm, yshift=-0.03cm, line width=1.1pt] (6-0.001,4+0.001) grid (6.5-0.0001,2.5-0.001);
        \path [anchor = center, fill=Melon, opacity=0.2, xshift=0.05cm, yshift=-0.03cm] (6,4) rectangle (6.5,2.5);
    \end{scope}

    \node () at (5.95,3) {\small{...}};

    \draw[decorate, decoration={brace,mirror}, line width=0.3mm] (4.4,2.1) -- (7.5,2.1);
    \node[align=center] () at (6,1.5) {\small{$\in \mathbb{R}^{n \times d_{AE}}$}};

    \node (Melon_sample_masked) at (3.65,3.5) {};
    \node (Melon_sample_masked_right) at (4.0,3.5) {};
    
    
    \begin{scope}[]
        \draw[Thistle, step=0.5cm, thick, xshift=0.05cm, yshift=-0.03cm, line width=1.1pt] (4.5-0.001,0.5+0.001) grid (5-0.0001,-2.5-0.001);
        \path [anchor = center, fill=Thistle, opacity=0.2, xshift=0.05cm, yshift=-0.03cm] (4.5,0.5) rectangle (5,-2.5);
    \end{scope}

    \begin{scope}[xshift=0.1cm,]
        \draw[Thistle, step=0.5cm, thick, xshift=0.05cm, yshift=-0.03cm, line width=1.1pt] (5-0.001,0.5+0.001) grid (5.5-0.0001,-2.5-0.001);
        \path [anchor = center, fill=Thistle, opacity=0.2, xshift=0.05cm, yshift=-0.03cm] (5,0.5) rectangle (5.5,-2.5);
    \end{scope}

    \begin{scope}[xshift=0.7cm,]
        \draw[Thistle, step=0.5cm, thick, xshift=0.05cm, yshift=-0.03cm, line width=1.1pt] (5.5-0.001,0.5+0.001) grid (6-0.0001,-2.5-0.001);
        \path [anchor = center, fill=Thistle, opacity=0.2, xshift=0.05cm, yshift=-0.03cm] (5.5,0.5) rectangle (6,-2.5);
    \end{scope}

    \begin{scope}[xshift=0.8cm,]
        \draw[Thistle, step=0.5cm, thick, xshift=0.05cm, yshift=-0.03cm, line width=1.1pt] (6-0.001,0.5+0.001) grid (6.5-0.0001,-2.5-0.001);
        \path [anchor = center, fill=Thistle, opacity=0.2, xshift=0.05cm, yshift=-0.03cm] (6,0.5) rectangle (6.5,-2.5);
    \end{scope}

    \node () at (5.95,-1) {\small{...}};
    
    \draw[decorate, decoration={brace,mirror}, line width=0.3mm] (4.4,-2.6) -- (7.5,-2.6);
    \node[align=center] () at (6,-3.1) {\small{$\in \mathbb{R}^{n_p \times d_{LLM}}$}};

    
    \node (Thistle_sample_left) at (4.65, -1.) {};
    \node (Thistle_sample_right) at (7.65,-1.) {};
    
    \node (Melon_sample_z_left) at (4.65,3) {};
    \node (Melon_sample_z_right) at (7.15,3) {};

    \draw[->, line width=0.3mm] (f_AE.east) ++ (0.1,0) -- (Melon_sample_z_left.west);
    \draw[->, line width=0.3mm] (f_emb.east) ++ (0.1,0) -- (Thistle_sample_left.west);

    \draw[->, line width=0.3mm] (prompt.east) ++ (-0.1,0) -- (f_emb.west);
    \draw[->, line width=0.3mm] (audio.east) ++ (0,0) -- (f_AE.west);


    \node[rectangle, draw, fill=cyan, minimum width=1.cm, minimum height=1.1cm, align=center, rounded corners, opacity=0.3, line width=1.25pt] (downsampling_g) at (9.35,3) {};
    \node () at (9.35,3) {$g$};
    \node[align=center] () at (9.35, 4.9) {\small{Downsampling} \\[-0.5ex] \small{Module}};

    \draw[->, line width=0.3mm] (Melon_sample_z_right.east) ++ (0.1,0) -- (downsampling_g.west);


    \begin{scope}[xshift=0.5cm, yshift=4.05cm]
        \begin{scope}[]
            \draw[Melon, step=0.5cm, thick, xshift=0.05cm, yshift=-0.03cm, line width=1.1pt] (11-0.001,0.5+0.001) grid (11.5-0.0001,-2.5-0.001);
            \path [anchor = center, fill=Melon, opacity=0.2, xshift=0.05cm, yshift=-0.03cm] (11,0.5) rectangle (11.5,-2.5);
        \end{scope}
    
        \begin{scope}[xshift=0.1cm,]
            \draw[Melon, step=0.5cm, thick, xshift=0.05cm, yshift=-0.03cm, line width=1.1pt] (11.5-0.001,0.5+0.001) grid (12-0.0001,-2.5-0.001);
            \path [anchor = center, fill=Melon, opacity=0.2, xshift=0.05cm, yshift=-0.03cm] (11.5,0.5) rectangle (12,-2.5);
        \end{scope}
    
        \begin{scope}[xshift=0.7cm,]
            \draw[Melon, step=0.5cm, thick, xshift=0.05cm, yshift=-0.03cm, line width=1.1pt] (12-0.001,0.5+0.001) grid (12.5-0.0001,-2.5-0.001);
            \path [anchor = center, fill=Melon, opacity=0.2, xshift=0.05cm, yshift=-0.03cm] (12,0.5) rectangle (12.5,-2.5);
        \end{scope}
    
        \begin{scope}[xshift=0.8cm,]
            \draw[Melon, step=0.5cm, thick, xshift=0.05cm, yshift=-0.03cm, line width=1.1pt] (12.5-0.001,0.5+0.001) grid (13-0.0001,-2.5-0.001);
            \path [anchor = center, fill=Melon, opacity=0.2, xshift=0.05cm, yshift=-0.03cm] (12.5,0.5) rectangle (13,-2.5);
        \end{scope}
    
        \node (left_downsampled_audio) at (11.2,-1.05) {};
        \node (right_downsampled_audio) at (13.75,-1) {};
    
        \node () at (12.45,-1) {\small{...}};
        
        \draw[decorate, decoration={brace,mirror}, line width=0.3mm] (10.9,-2.6) -- (14,-2.6);
        \node[align=center] () at (12.5,-3.1) {\small{$\in \mathbb{R}^{n_q \times d_{LLM}}$}};
    \end{scope}

    \draw[->, line width=0.3mm] (downsampling_g.east) ++ (0.05,0) -- (left_downsampled_audio.west);

    \node[rectangle, draw, fill=emobotyellow, minimum width=1.cm, minimum height=1.5cm, align=center, rounded corners, fill opacity=0.7, draw opacity=0.3, line width=1.5pt] (f_llm) at (19,1) {};
    \node () at (19, 1) {$f_{LLM}$};

    \node[rectangle, draw, minimum width=1.2cm, minimum height=0.5cm, align=center, rounded corners, opacity=0.4, line width=1.5pt] (f_emb) at (16,1) {};
    \node (concat) at (16, 1) {\small{$\texttt{Concat}$}};

    \draw[->, rounded corners=6pt, line width=0.3mm] (Thistle_sample_right.west) -- ++ (8.55,0) -- (concat.south);
    \draw[->, rounded corners=6pt, line width=0.3mm] (right_downsampled_audio.east) -- ++ (1.5,0) -- (concat.north);
    \draw[->, line width=0.3mm] (concat.east) -- (f_llm.west);


    \node (y) at (21.5, 1) {$\hat{\mathbf{y}}$};

    \draw[->, line width=0.3mm] (f_llm.east) ++ (0.1,0) -- (y.west);

    \node[align=center] () at (3.9,3.4) {\small{\color{burgundy}{(a)}}};
    \node[align=center] () at (3.9,-0.65) {\small{\color{burgundy}{(b)}}};
    \node[align=center] () at (10.85,3.5) {\small{\color{burgundy}{(c)}}};
    \node[align=center] () at (17.5,1.5) {\small{\color{burgundy}{(d)}}};

\end{tikzpicture}

%% file: 4_experiments.tex
\section{Experiments}
\label{sec:results}

    \subsection{Experimental settings}

        \paragraph{Dataset} For ASR training, we rely on the Librispeech dataset \cite{librispeech} as well as MSP-Podcast \cite{msp-podcast} since the transcript is also provided. Regarding SER, we only rely on the MSP-Podcast dataset for training. We evaluate the ability of \EmoLLM on the SER task on the \texttt{test1} share of MSP-Podcast.
        
        \paragraph{Training settings} We use AdamW \cite{loshchilov2018decoupled} as the optimizer with learning rate $5\cdot10^{-4}$ and weight decay 0.01. We also rely on linear scheduling with a warm-up on 10\% of the phase's training steps. After the downsampling module warm-up phase, the scheduling is re-started when the LoRA adapters are added to the LLM. We use WavLM \cite{wavLM} as the audio feature extractor and Llama 3.2-3B-Instruct \cite{grattafiori2024llama3herdmodels} as the foundation language model. The LoRA adapters are added to the following layers of the LLM: [\texttt{q\_proj}, \texttt{v\_proj}, \texttt{k\_proj}, \texttt{o\_proj}, \texttt{gate\_proj}, \texttt{up\_proj}, \texttt{down\_proj}], while contrary to \cite{Das2024SpeechVerseAL} we do not add LoRA adapters to the audio encoder and keep it frozen. LoRA adapter's rank is set to 8, with dropout $0.1$ and $\alpha$ equal to 16. For the downsampling module, implemented as a QPMapper, we use 32 learnable queries, 2 transformer layers with 8 attention heads each, and an embedding dimension of 768. The output of the downsampling module is then mapped to the dimension of the LLM using a learned linear layer.
        We set the effective batch size to 512 for all three phases. Epochs for phase \textcolor{emobotyellow}{\bf P1} and \textcolor{emobotyellow}{\bf P2} are set to 10, while phase \textcolor{emobotyellow}{\bf P3} lasts for a maximum of 20 epochs. We rely on early stopping on the validation SER loss, and stop after two consecutive epochs without improvement. To enhance the performance of \EmoLLM for emotion prediction, we progressively decrease the weight of the ASR task in the optimized loss. For that purpose,  we rely on a simple linear scheduler that attributes an equal weight to both ASR and Emotion Recognition losses for the first epoch of \textcolor{emobotyellow}{\bf P3} while progressively decreasing the weight of the ASR loss to 0 for the last epoch.

        \begin{figure}[t!]
            \centering
            \includegraphics[width=0.95\linewidth]{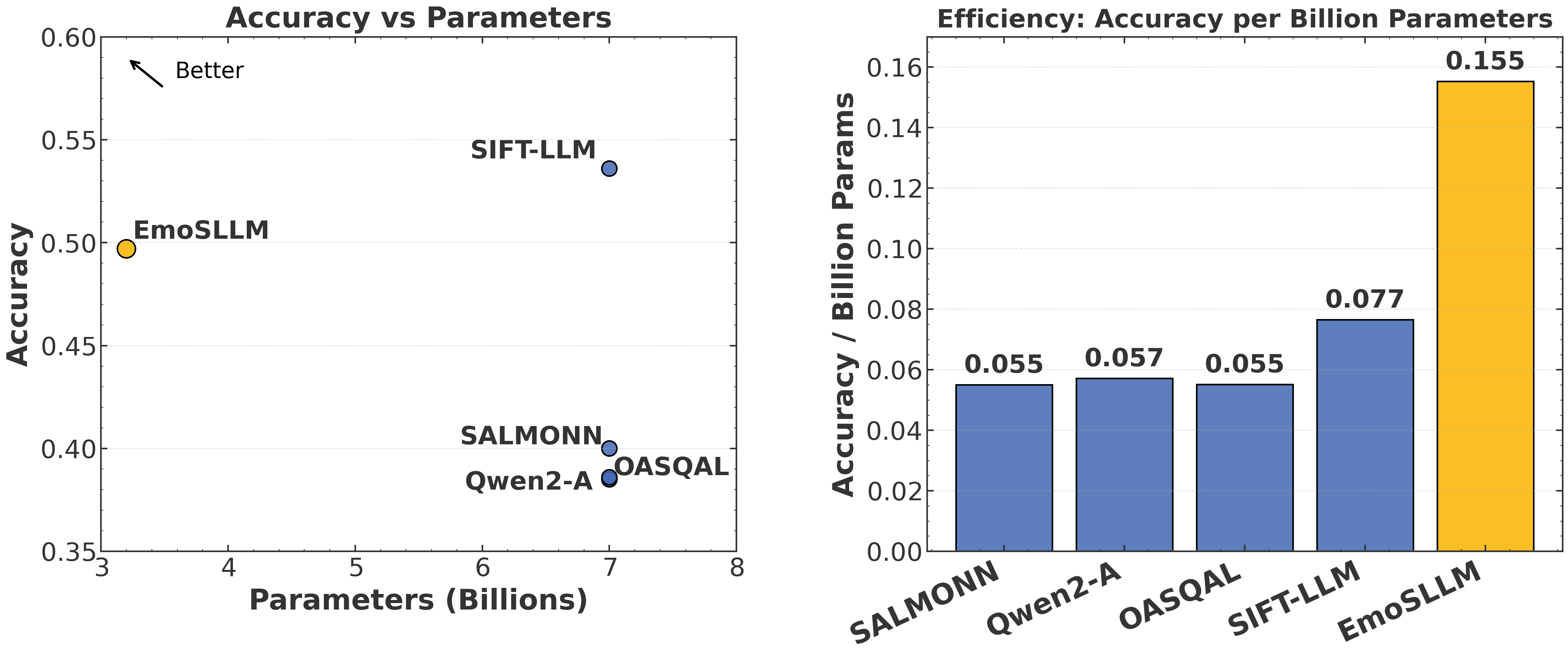}
            \caption{\textbf{Performance Comparison}. Performance comparison with existing Audio-Language Models that perform speech emotion recognition, Qwen2-Audio-Instruct \cite{Qwen-Audio} (Qwen2-A), OASQA-LLM \cite{pandey2025sift50mlargescalemultilingualdataset} (OASQAL), and SIFT-LLM \cite{pandey2025sift50mlargescalemultilingualdataset}. We compare \EmoLLM with existing methods on MSP-Podcast test1 and observe that with fewer parameters and significantly less training time, our model displays strong SER performance as it outperforms all methods except SIFT-LLM.}
            \label{fig:perf}
        \end{figure}

        \paragraph{Prompt setting} As further discussed in Section \ref{subsec:add_features}, we refer to \EmoLLM as the variant trained with additional paralinguistic features and 1-shot example hinting provided in $\z$. In contrast, \EmoLLM-base denotes the baseline architecture trained with standard prompts and without any auxiliary information in $\z$, as detailed in Sections \ref{app:subsec:asr}, \ref{app:subsec:ser}, and \ref{app:subsec:joint_decoding}. Both models are trained using identical hyperparameters. Also, for both \EmoLLM and \EmoLLM-base, inference is performed using joint decoding as detailed in section \ref{subsec:joint_prediction}, where the model has access to the transcript when forming its prediction. We discuss the added value of including in $\z$ additional paralinguistic features and $n$-shot example hinting in section \ref{sec:discussion}.

        \paragraph{Compute} Overall, including LoRA layers and the downsampling module, our model has 12.5M trainable parameters and 3.2B parameters overall. Our model is trained on 16 H100 Nvidia GPUs with PyTorch Lightning\footnote{\url{https://lightning.ai/}}.

       \paragraph{Benchmark} To ensure a rigorous evaluation, we benchmark our approach against existing Speech-Text LLMs that incorporate SER capabilities. For instance, we compare to SALMONN-7B \cite{tang2024salmonn}, Qwen2-Audio-7B-Instruct \cite{Qwen-Audio}, OASQA-LLM \cite{pandey2025sift50mlargescalemultilingualdataset} and SIFT-LLM \cite{pandey2025sift50mlargescalemultilingualdataset} using the test1 share of MSP-Podcast and the unweighted accuracy metric following previous work \cite{Qwen-Audio, pandey2025sift50mlargescalemultilingualdataset, tang2024salmonn, pandey2025sift50mlargescalemultilingualdataset}. Although related to our work, we are unable to include Speechverse \cite{Das2024SpeechVerseAL} in our benchmark as their model is not publicly available, and they only provide their model's performance for SER on the 4-class classification problem on MSP-Podcast. Regarding the remaining models to which we compare \EmoLLM, we provide details on training hours in table \ref{tab:compute_hours} and observe that all models requires significantly more training hours than our proposed model.

    \subsection{Results}
        \label{subsec:results}
        Figure \ref{fig:perf} (left) illustrates the performance of \EmoLLM in relation to existing methods, plotted against their respective parameter counts.
        Our results demonstrate that \EmoLLM achieves notable performance despite its smaller number of parameters. \EmoLLM significantly outperforms SALMONN \cite{tang2024salmonn},  Qwen2-Audio-Instruct \cite{Qwen-Audio}, and OASQAL \cite{pandey2025sift50mlargescalemultilingualdataset}, while having less than half (3.2B) the number of parameters than competing methods (7B+). 
        
        While SIFT-LLM exhibits superior performance in emotion prediction, this advantage comes with substantially higher computational requirements, as highlighted in Table \ref{tab:compute_hours}. We attribute the performance gap between \EmoLLM and SIFT-LLM to two primary factors. First, the backbone LLM used in their approach, Qwen2.5-7B-instruct \cite{qwen2025qwen25technicalreport}, has a significantly larger number of parameters than ours. This parameter advantage likely provides SIFT-LLM with greater representational capacity and more robust contextual understanding. Second, SIFT-LLM's multi-task training regime exposes it to significantly more diverse data, enabling better cross-modal feature learning and more generalizable representations. This multi-task approach may create synergy where emotion recognition benefits from related speech understanding tasks. Despite these advantages of SIFT-LLM, it is noteworthy that \EmoLLM achieves competitive performance while maintaining a significantly smaller parameter footprint, suggesting greater computational efficiency as shown on Figure \ref{fig:perf} (right).

%% file: 5_discussion.tex
\section{Discussion}
\label{sec:discussion}

    \subsection{Large Language Model}
        \label{subsec:disc_llm}

            To investigate the impact of both the LLM's architecture and training strategy, as well as parameter count, we compare the performance obtained by \EmoLLM when using Qwen3-4B \cite{qwen3} and Llama 3.2-3B-Instruct \cite{grattafiori2024llama3herdmodels} as $f_{LLM}$.

    \subsection{Additional features}
        \label{subsec:add_features}

        Previous work \cite{10888591} has demonstrated that adding paralinguistic audio features may boost emotion recognition using text-only LLMs. 
        \cite{10888591} only rely on the audio transcript and curated prompts for emotion recognition, in which case including additional paralinguistic information in the audio logically boosts the performance for emotion recognition. In our case, some paralinguistic information is likely already contained in the audio tokens.
        \begin{wraptable}{r}{0.4\textwidth}
            \centering
            \caption{Performance comparison between performance with and without paralinguistic features in natural languages on test1 from MSP-Podcast \cite{msp-podcast}. }
            \label{tab:add_features}
            \begin{tabular}{cc}
                \toprule
                 Add. features & Accuracy ($\uparrow$)  \\
                 \midrule
                 $\times$ & $0.458$ \\
                 \checkmark & $\mathbf{0.469}$\\
                  \bottomrule
            \end{tabular}
        \end{wraptable}
        Nevertheless, we investigate whether adding this information in the form of textual tokens in $\z$ might enhance our model's performance.
        We include the following paralinguistic features: loudness, average pitch, pitch range, jitter and shimmer. Also, following \cite{10888591} we include in $\z$ the gender of the speaker. 
        Rather than directly providing the value of the features, we provide the binned paralinguistic features in three classes [\texttt{'low'}, \texttt{ 'medium'}, \texttt{ 'high'}] that each represent a third of the values based on the training set.
        \begin{wraptable}{l}{0.3\textwidth}
            \vspace{-.47cm}
            \centering
            \caption{Performance comparison between $n$-shot hinting on test1 from MSP-Podcast \cite{msp-podcast}. }
            \label{tab:n_shot}
            \begin{tabular}{cc}
                \toprule
                 Hint  & Accuracy ($\uparrow$)  \\
                 \midrule
                  $0$-shot & $0.458$ \\
                  $1$-shot & $0.473$ \\
                  $2$-shot & $\mathbf{0.474}$ \\
                  \bottomrule
            \end{tabular}
            \vspace{-.2cm}
        \end{wraptable}
        We include those tokens by sampling among 5 introductory sentences and randomizing the order in which the paralinguistic features are provided.
        We display in table \ref{tab:add_features}, the performance of \EmoLLM-base when trained with $\z$ including the additional features and with an empty $\z$. We observe that including paralinguistic features in natural language inputs improves the performance of emotion prediction, increasing accuracy from 45.8\% to 46.9\%, a gain of 1.1 percentage points.

        \paragraph{Few-shot format hinting}
         We also investigate whether providing during phase \textcolor{emobotyellow}{\bf P3}, $n$ examples of the expected output structure in the user prompt, might help enhance the performance of \EmoLLM-base. See appendix \ref{app:subsec:example} for an example of such prompt strategy. 
        We provide in table \ref{tab:n_shot} the performance of \EmoLLM-base trained with $\z$ including $n$ examples in the user prompt for $n \in [0,1,2]$. We chose to keep $n$ small as we expect the marginal gain to be quite small for higher values while increasing the computational cost. We observe a significant difference between \EmoLLM-base without any hint ($0$-shot) and its performance when enhanced with the $1$-shot and $2$-shot hinting strategies as they display a respective gain of 1.5 and 1.6 percentage points over the $0$-shot approach. Since $1$-shot and $2$-shot hinting provide similar performance, we chose to keep $1$-shot hinting in our main approach as it involves a lower computational cost.

    \subsection{Joint prediction}
        \label{subsec:joint_prediction}

        \paragraph{Training} During phase \textcolor{emobotyellow}{\bf P3} training, when the emotion is available for a sample, we provide a prompt that asks for joint prediction. In other words, the model is asked to perform simultaneously the ASR and SER task. We believe that this could only be beneficial as it ensures that the model uses both semantic, linguistic and paralinguistic features to form its emotion prediction. See appendix \ref{app:subsec:joint_decoding} for an illustrative example.
        
        \paragraph{Inference}
        \begin{wraptable}{l}{0.4\textwidth}
            \centering
            \vspace{-.4cm}
            \caption{\textbf{Prompt Strategies}. Performance comparison of different prompting strategies during inference on the test1 split of MSP-Podcast \cite{msp-podcast}. }
            \label{tab:joint_prediction}
            \begin{tabular}{lc}
                \toprule
                 Prompt Strategy & Accuracy ($\uparrow$)  \\
                 \midrule
                 \texttt{SER-only} & 0.417 \\
                 \texttt{Prompt-hint} & 0.431 \\
                 \EmoLLM & $\mathbf{0.497}$ \\
                 \bottomrule
            \end{tabular}
        \end{wraptable}
        The first approach, referred to as \texttt{SER-only}, involves prompting the LLM exclusively for the SER task without any auxiliary information. To assess the utility of providing transcript information as contextual hints, we explore two additional approaches.
        First, we consider providing the transcript in the user prompt, introduced by \hlbox[emobotyellow!40]{\texttt{"Use the following transcript to help you }} \hlbox[emobotyellow!40]{\texttt{ predict the emotion:"}}, we refer to this approach as \texttt{Prompt-hint}. Note that this approach was never used during the training phase. Second, we consider providing the same user prompt as the ones seen during training, but we provide the beginning of the answer to the LLM and ask it to complete it. In other words, we ask the LLM to perform auto-regressive generation where its context contains the user prompt followed by the beginning of the assistant's answer, \hlbox[emobotyellow!40]{\texttt{"| ASR: <transcript> | Emotion:"}}. We refer to this last approach as \EmoLLM. See section \ref{app:subsec:transcript_hinting} for examples of such prompts. Comparison of the performance between these approaches is displayed in Table \ref{tab:joint_prediction}.
        
        Overall, we find that incorporating the transcript into the LLM’s input significantly enhances SER accuracy. Both \texttt{Prompt-hint} and \EmoLLM outperform the \texttt{SER-only} baseline. However, providing the transcript within the assistant’s response, as done in \EmoLLM, proves more effective than embedding it in the user prompt. Specifically, \EmoLLM achieves an accuracy of 0.497, compared to 0.431 for \texttt{Prompt-hint}. The LLM’s unfamiliarity with user prompts containing transcripts, since it was not exposed to such prompts during training, likely contributes to this performance gap.

    \subsection{Audio Encoder}
        \label{subsec:audio_encoder}
        \begin{wraptable}{r}{0.38\textwidth}
            \vspace{-0.5cm}
            \centering
            \caption{\textbf{Audio Encoder}. Performance comparison between pretrained audio encoders in \EmoLLM on test1 from MSP-Podcast \cite{msp-podcast}.}
            \label{tab:audio_encoder}
            \begin{tabular}{lcc}
                \toprule
                Audio Encoder  & Accuracy ($\uparrow$)  \\
                \midrule
                wav2vec 2.0 & $0.471$ \\
                wavLM & $\mathbf{0.497}$\\
                \bottomrule
            \end{tabular}
            \vspace{-0.2cm}
        \end{wraptable}
        We assess the impact of the choice of backbone audio encoder by replacing WavLM \cite{wavLM} with Robust wav2vec 2.0 \cite{hsu21_interspeech}. The alternative model is trained using the same hyperparameters, training curriculum, and prompting strategy as the original configuration. Table \ref{tab:audio_encoder} compares the performance of \EmoLLM when using WavLM versus Robust wav2vec 2.0 as the pretrained audio encoder. While both encoders yield strong results, WavLM consistently outperforms Robust wav2vec 2.0 in this setup. However, it is important to note that the hyperparameters were optimized for WavLM and may not be ideal for wav2vec 2.0, potentially limiting the latter’s performance.

%% file: 6_conclusion.tex
\section{Conclusion}
\label{sec:conclusion}

    This paper introduced \EmoLLM, a novel and computationally efficient approach for speech emotion recognition (SER) that effectively integrates audio and text modalities using LLMs. Our experimental results on standard SER benchmarks demonstrate that \EmoLLM outperforms most existing Speech-Text LLMs in the literature while requiring significantly fewer parameters and less training time. This highlights \EmoLLM's effectiveness and paves the way for more efficient and privacy-preserving applications in areas like human-computer interaction and mental health monitoring. 

    \paragraph{Limitations and future work} While \EmoLLM shows strong performance and efficiency, it is still surpassed by SIFT-LLM. This is likely due to SIFT-LLM benefiting from a larger backbone LLM and exposure to a significantly greater volume of multi-task training data. This suggests that even with parameter efficiency, the scale of the base LLM and training data diversity remain crucial. Furthermore, achieving true on-device deployment for multimodal LLMs still requires substantially reducing the overall parameter count. Future work could explore the integration of smaller backbone LLMs, model compression techniques such as quantization, or extending \EmoLLM to handle a broader range of multimodal inputs.

\section*{Acknowledgement}

    This work was granted access to the HPC resources of IDRIS under the allocation 2025-AD010616049 made by GENCI.

%% file: appendix.tex
\section{Prompts}
\label{app:sec:example_prompts}

    We provide in this section a more comprehensive description of the prompt structures used to train \EmoLLM.

    \subsection{System prompt}
     \label{app:subsec:system}
     We carefully design a system that details to the LLM the task at hand, while providing useful information about the expected input and output structures. We provide hereafter a snippet of the curated system prompt.

     \begin{block}{System prompt}
            \begin{tabbing}
            \hspace{0.5cm}\=\hspace{1cm}\=\kill
                \textbf{\{}\\
                \textcolor{MidnightBlue}{\textbf{"role"}}:\\
                \> \textcolor{black}{"system"},\\
                \textcolor{MidnightBlue}{\textbf{"content"}}:\\
                \>"You are a highly capable assistant specialized in audio processing tasks. \\
                \>You receive inputs containing audio token representations followed by text \\
                \>instructions, and return structured answer. \\ \\
                \>You may be asked to perform: \\
                \>\quad1. **Automatic Speech Recognition (ASR)** — transcribe the spoken content.\\
                \>\quad2. **Speech Emotion Recognition (SER)** — identify the emotion expressed\\
                \>in the audio.\\ \\
                \>Follow one of the two output formats:\\
                \>- For ASR-only tasks:\\
                \>\quad\quad '| ASR: <transcription> |'\\
                \>- For SER-only tasks:\\
                \>\quad\quad'| Emotion: <emotion code> |'\\
                \>For tasks involving both ASR and SER, use the following format:\\
                \>\quad\quad'| ASR: <transcription> | Emotion: <emotion code> |'\\ \\
                \>Emotion must be provided as a single letter chosen from the following emotion \\
                \>codes:\\
                \>\quad- A: Angry \\
                \>\quad- S: Sad \\
                \>\quad- H: Happy \\
                \>\quad- U: Surprise \\
                \>\quad- F: Fear \\
                \>\quad- D: Disgust \\
                \>\quad- C: Contempt \\
                \>\quad- N: Neutral \\
                \>\quad- O: Other \\
                \>\texttt{(...)"} \\
                \textbf{\}}
            \end{tabbing}
        \end{block}

    \subsection{Automatic Speech Recognition (ASR) prompt}
    \label{app:subsec:asr}
        As previously discussed in the main section of the paper, for each sample we select a prompt among a curated selection of prompts detailing the expected task at hand. We provide hereafter an example an ASR prompt used during training.

        \begin{block}{ASR prompt}
            \begin{tabbing}
            \hspace{0.5cm}\=\hspace{1cm}\=\kill
                \textbf{\{}\\
                \textcolor{MidnightBlue}{\textbf{"role"}}:\\
                \>\textcolor{black}{"user"},\\
                \textcolor{MidnightBlue}{\textbf{"content"}}:\\
                \>\textcolor{black}{"You will now perform the following audio-based task.} \\
                \>\textcolor{black}{Task: **Automatic Speech Recognition (ASR)**.}\\
                \>\textcolor{black}{Transcribe the preceding audio into written text."} \\
                \textbf{\}}
            \end{tabbing}
        \end{block}

    \subsection{Speech Emotion Recognition (SER) prompt}
    \label{app:subsec:ser}
        We provide hereafter an example of a vanilla SER prompt used during training.

        \begin{block}{SER prompt}
            \begin{tabbing}
            \hspace{0.5cm}\=\hspace{1cm}\=\kill
                \textbf{\{}\\
                \textcolor{MidnightBlue}{\textbf{"role"}}:\\
                \>\textcolor{black}{"user"},\\
                \textcolor{MidnightBlue}{\textbf{"content"}}:\\
                \>\textcolor{black}{"You will now perform the following audio-based task.} \\
                \>\textcolor{black}{Task: **Speech Emotion Recognition**.}\\
                \>\textcolor{black}{Classify the tone of the speaker in the preceding audio."}\\
                \textbf{\}}
            \end{tabbing}
        \end{block}

    \subsection{Joint decoding prompt}
    \label{app:subsec:joint_decoding}
        We provide hereafter an example of a vanilla joint decoding prompt used during training.

        \begin{block}{Joint decoding prompt}
            \begin{tabbing}
            \hspace{0.5cm}\=\hspace{1cm}\=\kill
                \textbf{\{}\\
                \textcolor{MidnightBlue}{\textbf{"role"}}:\\
                \>\textcolor{black}{"user"},\\
                \textcolor{MidnightBlue}{\textbf{"content"}}:\\
                \>\textcolor{black}{"Perform the following audio-based tasks in the order as described.} \\
                \>\quad\textcolor{black}{1. Task: **Automatic Speech Recognition (ASR)**.}\\
                \>\quad\textcolor{black}{Identify and write down the words spoken in the preceding audio.} \\
                \>\quad\textcolor{black}{2. Task: **Speech Emotion Recognition**.}\\
                \>\quad\textcolor{black}{Analyze the audio and determine the emotional state of the speaker."} \\
                \textbf{\}}
            \end{tabbing}
        \end{block}

    \subsection{Supplementary features hinting}
    \label{app:subsec:supp_features}

        We provide a variety of supplementary features to guide the LLM in its prediction. We include features ranging from the gender of the person speaking, to some paralinguistic features. Supplementary feature can be combined or taken in isolation. We hereafter provide an example of how the supplementary features are included in the prompts as shown in sections \ref{app:subsec:asr}, \ref{app:subsec:ser} and \ref{app:subsec:joint_decoding}.

       \begin{block}{Supplementary feature prompt}
            \begin{tabbing}
            \hspace{0.5cm}\=\hspace{1cm}\=\kill
                \textbf{\{}\\
                \textcolor{MidnightBlue}{\textbf{"role"}}:\\
                \>\textcolor{black}{"user"},\\
                \textcolor{MidnightBlue}{\textbf{"content"}}:\\
                \>"\texttt{
                (...)} \\
                \>The speaker in this audio is \texttt{\$gender}. \\
                \>\textcolor{black}{Here's a breakdown of paralinguistic cues in the audio: }\\
                \>\quad\textcolor{black}{- loudness: 'low'} \\
                \>\quad\textcolor{black}{- pitch: 'medium'} \\
                \>\quad\textcolor{black}{- pitch range: 'high'} \\
                \>\quad\textcolor{black}{- jitter: 'low'} \\
                \>\quad\textcolor{black}{- shimmer: 'high'} \\
                \>\quad\textcolor{black}{\ttfamily (...)"}\\
                \textbf{\}}
            \end{tabbing}
        \end{block}

    \subsection{Example hinting}
    \label{app:subsec:example}

        We also include the possibility of including $n$ examples of answers in the format we are expecting. We include hereafter an example for example hinting in the case of joint decoding. Note that we also include this for SER-only and ASR-only prompts.

       \begin{block}{Few shot example prompt}
            \begin{tabbing}
            \hspace{0.5cm}\=\hspace{1cm}\=\kill
                \textbf{\{}\\
                \textcolor{MidnightBlue}{\textbf{"role"}}:\\
                \>\textcolor{black}{"user"},\\
                \textcolor{MidnightBlue}{\textbf{"content"}}:\\
                \>\textcolor{black}{\ttfamily "(...)}\\
                \>\textcolor{black}{Here are some examples of the expected output:} \\
                \>\quad\textcolor{black}{- '| ASR: You’re such a disappointment. | Emotion: C |'} \\
                \>\quad\textcolor{black}{-  '| ASR: I'm speechless... didn't expect that. | Emotion: U |',} \\
                \>\quad\textcolor{black}{- '| ASR: I can’t shake this feeling of dread. | Emotion: F |'"} \\
                \>\textcolor{black}{\ttfamily (...)"}\\
                \textbf{\}}
            \end{tabbing}
        \end{block}

    \subsection{Transcript hinting}
    \label{app:subsec:transcript_hinting}

        As mentioned in section \ref{subsec:joint_prediction}, in inference we consider two alternatives to include the transcript as a hint to guide the LLM to make its prediction on the emotion class.

        \paragraph{Prompt-hint} First, we consider including the transcript in the user prompt with an introductory sentence. This approach is never used in training.

            \begin{block}{\texttt{Prompt-hint} prompt}
            \begin{tabbing}
            \hspace{0.5cm}\=\hspace{1cm}\=\kill
                \textbf{\{}\\
                \textcolor{MidnightBlue}{\textbf{"role"}}:\\
                \>\textcolor{black}{"user"},\\
                \textcolor{MidnightBlue}{\textbf{"content"}}:\\
                \>\textcolor{black}{"\texttt{(...)}} \\
                \>\textcolor{black}{Use the following transcript to help you classify the emotion:} \\
                \>\textcolor{black}{\texttt{<transcript>}}\\
                \>\textcolor{black}{\texttt{(...)}"}\\
                \textbf{\}}
            \end{tabbing}
        \end{block}

        \paragraph{\EmoLLMbf} For our main pipeline, we consider an alternative that most resembles the task that the LLM performs in the training phase. We use as a user prompt the joint-decoding prompt as detailed in section \ref{app:subsec:joint_decoding}, and ask the LLM to auto-regressively generate tokens given the first part of the answer that contains the true transcript.

            \begin{block}{EmoSLLM prompt}
            \begin{tabbing}
            \hspace{0.5cm}\=\hspace{1cm}\=\kill
                \textbf{\{}\\
                \textcolor{MidnightBlue}{\textbf{"role"}}:\\
                \>\textcolor{black}{"user"},\\
                \textcolor{MidnightBlue}{\textbf{"content"}}:\\
                \>\textcolor{black}{\texttt{<Joint decoding prompt>}} \\
                \textbf{\}},\\
                \textbf{\{}\\
                \textcolor{MidnightBlue}{\textbf{"role"}}:\\
                \>\textcolor{black}{"assistant"},\\
                \textcolor{MidnightBlue}{\textbf{"content"}}:\\
                \>\textcolor{black}{"| ASR: \texttt{<transcript>} | Emotion:"} \\
                \textbf{\}}
            \end{tabbing}
            \end{block}

    \section{Compute}
        \label{app:sec:compute}
        
        We display in table \ref{tab:compute_hours} the compute hours required to train the different models to which we compare our method. The displayed hours are directly extracted from the original papers.
        
        \begin{table}[h!]
            \centering
            \caption{Comparison of the total training hours and training hours specific to SER.}
            \begin{tabular}{ccc}
                \toprule
                 Model & Total training hours & SER training hours  \\
                 \midrule 
                 SALMONN \cite{tang2024salmonn} & $\sim 4400$ & $5$ \\ 
                 Qwen2-Audio-7B-Instruct \cite{Qwen-Audio} & $\sim 146500$ & $\sim 1000$\\
                 SIFT-LLM \cite{pandey2025sift50mlargescalemultilingualdataset} & $173483$ & $237$ \\
                 \midrule
                 \EmoLLM (Ours) & $\sim 320$ & $\sim 180$  \\
                 \bottomrule
            \end{tabular}
            \label{tab:compute_hours}
        \end{table}